\documentclass[aps,prd,showpacs,nofootinbib,floats,floatfix,preprintnumbers,groupedaddress,twocolumn]{revtex4}

\usepackage{bm}
\usepackage{latexsym}
\usepackage{dcolumn}
\usepackage{amsmath,amsfonts,amssymb}
\usepackage{graphicx,epsfig}
\usepackage{amsmath}
\usepackage{fancyhdr}
\usepackage{hyperref}
\usepackage{graphicx,epstopdf}
\hypersetup{
	colorlinks   = true, 
	urlcolor     = blue, 
	linkcolor    = blue, 
	citecolor   = red 
}

\def\p{\partial}

\begin{document}
\title{Thermogeometric phase transition in a unified framework}
\author{Rabin Banerjee$^{a}$\footnote {\color{blue} rabin@bose.res.in}}
\author{Bibhas Ranjan Majhi$^{b}$\footnote {\color{blue} bibhas.majhi@iitg.ernet.in}}
\author{Saurav Samanta$^{c}$\footnote {\color{blue} srvsmnt@gmail.com}}

\affiliation{$^a$S. N. Bose National Centre for Basic Sciences, JD Block, Sector III, Salt Lake, Kolkata-700098, India\\
$^b$Department of Physics, Indian Institute of Technology Guwahati, Guwahati 781039, Assam, India\\
$^c$Department of Physics, Narasinha Dutt College, 129, Belilious Road, Howrah 711101, India
}

\date{\today}

\begin{abstract}
Using geomterothermodynamics (GTD), we investigate the phase transition of black hole in a metric independent way. We show that for any black hole, curvature scalar (of equilibrium state space geometry) is singular at the point where specific heat diverges. Previously such a result could only be shown by taking specific examples on a case by case basis. A different type of phase transition, where inverse specific heat diverges, is also studied within this framework. We show that in the latter case, metric (of equilibrium state space geometry) is singular instead of curvature scalar. Since a metric singularity may be a coordinate artifact, we propose that GTD indicates that it is the singularity of specific heat and not inverse specific heat which indicates a phase transition of black holes.

\end{abstract}

\maketitle

{\section{Introduction}}
   Phase transition of usual thermodynamic systems like hydrodynamic, magnetic etc. is a thoroughly studied subject in physics. Gibbs introduced a thermodynamic potential which can describe the behaviour of such systems at the point of phase transition. He and Caratheodory also introduced a geometric method to study thermodynamics. Later Weinhold \cite{Weinhold} gave a firm basis of this approach by introducing a metric defined as the Hessian of the internal energy.

   A slightly different approach was taken by Ruppeiner \cite{Ruppeiner1}-\cite{Ruppeiner2} where Hessian of the entropy was taken as metric. In fact the Ruppeiner and Weinhold  metrics are conformaly related where inverse temperature is the conformal factor.
It has been proposed that physics of phase transition can be described by this metric formulation of thermodynamics. A curvature singularity typically indicates a phase transition and thus curvature is interpreted as a measure of thermodynamic interaction. This claim has been verified for various systems. For example, both Weinhold and Ruppeiner curvature diverge at the critical point of van der Waals system. 

The geometrical approach was also tested for different black hole metrics by pursuing a case by case study  \cite{Cai:1998ep} -- \cite{Zangeneh:2016fhy}. It has been observed that such a study is quite laborious but the final conclusion is identical -- the Ricci scalar diverges at the point where specific heat diverges and hence divergence of curvature is a signature of  phase transition.  
Although the above statement is almost universal,  there exists an uncomfortable feature. One finds that for black hole systems things are quite complicated. For Reissner-Nordstrom black hole, Ruppeiner metric is flat but Weinhold curvature diverges at the extremal limit. There are attempts to rectify the issue \cite{Shen:2005nu,Mirza:2007ev} but they are not self-sufficient.

 In this paper we develop a unified approach for studying phase transition in the geometrothermodynamic (GTD) approach. Not only do we dispense with a case by case investigation, but the disagreement between the Ruppeiner and Weinhold approaches is also avioded. The idea is since it is expected that the first law of thermodynamics is a universal property of the black holes, so the thermodynamical metric derivable from it is also a universal feature. If this is so, then  phase transition can be easily studied by looking at the Ricci scalar of this metric,  without considering any specific black hole metric. It will be shown in this paper that such an analysis is indeed possible. Here the metric is constructed by using the Legendre invariant formulation proposed in \cite{Alvarez:2008wa} (This is the rectified version of earlier proposal \cite{Quevedo:2006xk,Quevedo:2007mj}). Since the origin of disagreement, if any,  between Ruppeiner and Weinhold approaches is contained in their lack of Legendre invariance, this issue is naturally bypassed in our formalism. We find that, divergence of heat capacity means singularity of Legendre invariant metric curvature for any black hole. This result is quite important. It demonstrates the power of GTD to study the phase transition of black holes. The fact that a universal scheme could be developed for GTD is somewhat inspired by  recent papers \cite{Mandal:2016anc,Majhi:2016txt}, where two of the present authors, in a collaborative effort, presented a general scheme for discussing phase transition, along more conventional lines, in black holes without referring to any specific metric. In this context it may be recalled that besides GTD, there are other ways of looking at phase transition in black holes. These include, but are not restricted to, checking the divergence of specific heat and the inverse of isothermal compressibility \cite{Banerjee1,Majhi,Lala:2012jp}, AdS black holes \cite{Kubiznak:2012wp} (For recent reviews, see \cite{Altamirano:2014tva,Kubiznak:2016qmn}), Ehrenfest-like schemes \cite{Banerjee:2010qk}, apart from the original work of Davies \cite{Davies:1978zz} and later, Hawking and Page \cite{Hawking:1982dh}. 
 
 We have also looked at  at another type of phase transition. In some papers \cite{Kaburaki,Cai:1996df} it has been argued that phase transition of black hole is associated with a singularity of inverse heat capacity and moment of inertia. Now the question is: Can it be possible to rephrase this  in terms of thermo-geometry? Here we show that this means a singularity of equilibrium state space metric, instead of the curvature scalar. We then argue that since a singularity of metric may be due to the ``choice of coordinates'', it seems that GTD suggests that a 
  phase transition should be connected with curvature singularity i.e. divergence of heat capacity as was originally suggested by Davies.

\vskip 2mm
\noindent
{\section{Setup: a unified framework}}
To discuss the phase transition in terms of geometry, let us first construct the thermodynamical metric in a specific black hole metric independent way. The only condition that has to be imposed is the Legendre invariance of the metric. The staring point is the first law of black hole thermodynamics:
\begin{eqnarray}
TdS=dM-\sum_iY_idX_i
\label{TdS}
\end{eqnarray}
where $Y_i$ are generalized forces like electric potential ($\Phi$), angular velocity ($\Omega$) etc. and $X_i$ are generalized displacements like electric charge ($Q$), angular momentum ($J$) etc. Also, if there is any other hair in the black hole, that will be represented by the last term of the above relation. It says that entropy can be taken as a function of $M$ and $X_i$. In that case we can write
\begin{eqnarray}
dS & = & \left(\frac{\p S}{\p M}\right)_{X_i}dM+\sum_{i \ne j}\left(\frac{\p S}{\p X_i}\right)_{M,X_j}dX_i\\
& \equiv & S_M dM+ S_{X_i}dX_i
\label{dS}
\end{eqnarray}
Comparing the above equation with (\ref{TdS}), we find
\begin{eqnarray}
S_M=\frac{1}{T} \  \ {\textrm{and}} \  \ S_{X_i}=-\frac{Y_i}{T}~.
\label{S}
\end{eqnarray}
These are very important relations as they will be used several times in the rest of the calculations. Moreover, since the first law is a universal fact, the above ones are also very much universal by nature; i.e. they are not for any particular black hole metric. This is a very important fact and this will help us to give a general formulation of the GTD type phase transition.

   Now let us choose the thermodynamical coordinates as $\mathcal{Z}^A = (S,\mathcal{E}^a,\mathcal{I}^a)$ with $\mathcal{E}^a=(M,X_i)$ and $\mathcal{I}^a=(1/T,-Y_i/T)$. This is usually called as $S$-representation and here the fundamental one-form is given by $\Theta = dS-dM/T+(\sum_i Y_idX_i)/T$. The Legendre transformations can be taken as 
   \begin{eqnarray}
   &&(S,\mathcal{E}^a,\mathcal{I}^a)\rightarrow(\tilde{S},{\tilde{\mathcal{E}}}^a,{\tilde{\mathcal{I}}}^a)~;
   \nonumber
   \\
   &&S=\tilde{S}-\delta_{ab}\tilde{\mathcal{E}}^a\tilde{\mathcal{I}}^b, \,\,\,\ \mathcal{E}^a=-\tilde{\mathcal{I}}^a, \,\,\,\ \mathcal{I}^a=\tilde{\mathcal{E}}^a
   \end{eqnarray}
Within  this setup, following \cite{Alvarez:2008wa} one writes the Legendre invariant thermo-geometric metric as
\begin{equation}
g=\Theta^2+\left(\frac{M}{T}-\frac{\sum_iX_iY_i}{T}\right)\left[dMd\Big(\frac{1}{T}\Big)
\\
+dX_id\left(\frac{Y_i}{T}\right)\right].
\end{equation}
Then using (\ref{S}) and the fact that both $S_M$ and $S_{X_i}$ functions of ($M,X_i$) the above yields
\begin{equation}
g = \left(MS_M+\sum_iX_iS_{X_i}\right)\left[S_{MM}dM^2
- \sum_{i,j}S_{X_iX_j}dX_idX_j\right]
\label{metric}
\end{equation}
where $S_{MM}=(\p^2 S/ \p M^2)_{X_i}$ and $S_{X_iX_j}=(\p^2 S/\p X_i \p X_j)_{M}$. In getting the above relation we also used the first law of thermodynamics (\ref{TdS}). The last form of the metric is usually known as the induced metric. Note that it has been obtained upon using the first law. In the ensuing discussion  we shall work with this induced metric. 
\vskip 2mm
\noindent
\section{Phase transition}
  The geometrothermodynamic phase transition, as mentioned earlier, is mainly related to the Davies type phase transition. Here one looks for the divergence of the specific heat, defined by
  \begin{eqnarray}
C_{X_i}=T\left(\frac{\p S}{\p T}\right)_{X_i}=\left(\frac{\p M}{\p T}\right)_{X_i}~.
\end{eqnarray}
The point where the above quantity diverges, is the critical point. 
It is important to note that this can be expressed in terms of the $dM^2$ metric coefficient of (\ref{metric}).
Since
\begin{eqnarray}
S_{MM}\equiv \left(\frac{\p^2 S}{\p M^2}\right)_{X_i}=\frac{\p}{\p M}\left(\frac{1}{T}\right)_{X_i}=-\frac{1}{T^2}\left(\frac{\p T}{\p M}\right)_{X_i}~,
\end{eqnarray}
the specific heat takes the form:
\begin{eqnarray}
C_{X_i}=-\frac{1}{T^2S_{MM}}~.
\end{eqnarray}
It tells that for a non-extremal (i.e. $T\neq 0$) black hole, the critical point (i.e. $C_{X_i}\rightarrow\infty$) corresponds to the vanishing of $S_{MM}$. Therefore now our task is to calculate the Ricci scalar $R$ for the metric (\ref{metric}) and see if $R$ also diverges when $S_{MM}$ vanishes.
  
For that note that the metric (\ref{metric}) is in the following form:
\begin{eqnarray}
g=fdM^2+g_{AB}dX^AdX^B~,
\end{eqnarray}
where $f=\left(MS_M+\sum_iX_iS_{X_i}\right)S_{MM}$ and $g_{AB}$ are the rest of the metric coefficients. Now if $R$ diverges when $S_{MM}$ vanishes, then the divergent part of the Ricci scalar has to be proportional to inverse powers of $f$. Therefore let us look for the divergent terms in $R$. A direct calculation shows that the maximum order of the divergent part of $R$ when $f=0$ is given by 
\begin{eqnarray}
R|_{{\textrm{max diver}}} & \sim & \frac{1}{2} \frac{g_{AB}}{f^2}(\p_t f)(\p_t g_{AB}) + \frac{g_{AB}}{f^2}(\p_A f)(\p_B f) \nonumber\\
& \sim & {\mathcal O} \left(\frac{1}{S_{MM}^2}\right)~.
\end{eqnarray}
Hence the singularity in $R$ implies a singularity in the specific heat which indicates a phase transition. One point may be worth mentioning. The above is so only if numerator does not vanish at this point, which can not be shown to be true in general. Most of the cases it has been found that the numerator does not vanish where $S_{MM}$ is zero (See \cite{Alvarez:2008wa} for some explicit examples). But there is no guarantee for all cases. It could happen that the numerator eliminates the singularity and the scalar behaves as $0/0$, then one has to take care of it by the L'Hospital rule.

   Let us now discuss an interesting observation. Since at the critical point $S_{MM}$ vanishes and $X_i$s are constant, the metric (\ref{metric}) becomes null. This is similar to the horizon of a usual black hole -- horizon is a null surface. This piece of observation can be very important and interesting as in the black hole case the null surface plays the main role for its interpretation as a thermodynamic system. But at this point we are not sure if this can give further insight into the geometric phase transition.

  We next consider another type of phase transition which is characterised by the divergence of the following quantities \cite{Kaburaki}:
$\bar{\chi}_1  \equiv (\p^2 S/\p M^2)_{X_i}\equiv S_{MM}$ and
$\left(\bar{\chi}_2\right)_{ij}  \equiv (\p^2 S/\p X_i \p X_j)_M \equiv S_{X_iX_j}$. Note that these are proportional to the metric coefficients of our thermodynamic metric (\ref{metric}). Then a natural question comes in our mind: Can we give a geometric description of such type of phase transition? Below, a comment will be made on this point. 

  We first find out when the above quantities will diverge. To see this, one expresses them in the following forms:
\begin{eqnarray}
&&\bar{\chi}_1=-\frac{1}{T^2 C_{X_i}}=-\frac{\beta^2}{ C_{X_i}}~,
\nonumber
\\
&&\left(\bar{\chi}_2\right)_{ij}=-\frac{\beta^2}{ (I_M)_{ij}}
\end{eqnarray}
where $(I_M)_{ij}$ is the moment of inertia tensor, defined as
$(I_M)_{ij}=\beta\left(\frac{\p X_i}{\p \mu_j}\right)_M \ {\textrm{with}} \ \mu_i=\beta Y_i$ and $\beta$ is the inverse of the temperature.
Clearly, if one interprets the phase transition as $\bar{\chi}_1,\left(\bar{\chi}_2\right)_{ij}\rightarrow \infty$, then at the critical point
either $T=\frac{1}{\beta}\rightarrow 0$ or $C_{X_i},(I_M)_{ij}\rightarrow 0$; i.e. in one case at the critical point there is a possibility that the black hole becomes extremal.  Whereas for non-extremal situation, according to some authors, at phase transition inverse heat capacity and inverse moment of inertia diverge. Now since $\bar{\chi}_1$ and $\left(\bar{\chi}_2\right)_{ij}$ are coefficients of thermodynamic metric (\ref{metric}), this type of phase transition implies divergence of the metric coefficients. Moreover in this case the Ricci scalar vanishes as $C_{X_i}=0$ implies divergence of $S_{MM}$.
But it must be remembered that divergence of the metric coefficient is not a covariant statement; rather it is a ``coordinate dependent'' statement. Therefore one can write the metric in  new ``coordinates'' where there is no divergence in the metric coefficients. Hence the geometric description of such a phase transition is ambiguous. This is also supported by the fact that divergences of $\bar{\chi}_1$ and $\bar{\chi}_2$ lead to a vanishing $R$ which obviously contradicts the geometrical approach where phase transition is charactersied by a divergent $R$.

\vskip 2mm
\noindent
\section{Conclusions}

Though geometric method of thermodynamics has a long history, it is not often used to study physical systems. Two of the most important versions of this approach were given by Weinhold and Ruppeiner. These two theories are very similar and generally give consistent results for usual thermodynamic systems like van der Waals gas. In fact the  curvature scalar found from Weinhold and Ruppeiner metrics are known to diverge at the critical points of usual systems indicated by a singularity in the specific heat. But for black holes, there is some confusion. The results not only disagree  among themselves but also with other approaches. For example, for Reissner Nordstrom black hole, the curvature scalar of only the Weinhold metric is singular. That singularity is also not consistent with the singularity of specific heat.

Recent findings \cite{Alvarez:2008wa,Quevedo:2006xk,Quevedo:2007mj} show that the above contradictions are rooted in the fact that  neither Weinhold nor Ruppeiner metric is Legendre invariant and so is not suitable to describe thermodynamics. For various black holes like Reissner Nordstrom or Kerr this has been shown on a case by case basis i.e. singularity of curvature scalar (of Legendre invariant metric) coincides with the singularity of specific heat. Based on the original works \cite{Alvarez:2008wa,Quevedo:2006xk}, the method of geometrothermodynamics (GTD), presented here, incorporates  Legendre invariance  in a natural way through the use of the first law of thermodynamics. Since the first law is universal, the results following from such an approach are also expected to be universal. The Legendre invariant metric was calculated in a metric independent way  so that the results are general and valid for any black hole. The singularity of curvature scalar of this metric  coincides with the singularity of specific heat.  This important conclusion fills a  crucial  gap in our understanding of black hole phase transition from the point of view of GTD.

We have found another useful result. There has been some debate regarding the actual critical point of black hole phase transition. Is it the point where specific heat diverges or the point where inverse specific heat diverges?  Our results show that in the first case curvature scalar (of equilibrium state space geometry) diverges and in the second case metric (of equilibrium state space geometry) diverges. Since a metric singularity may be a coordinate artifact, we conclude that GTD indicates that actual black hole phase transition occurs at the point where specific heat diverges. This is conceptually satisfactory because normal thermodynamic systems behave exactly in a similar manner.  
    
\vskip 4mm
{\section*{Acknowledgments}}
The research of one of the authors (BRM) is supported by a START-UP RESEARCH GRANT (No. SG/PHY/P/BRM/01) from Indian Institute of Technology
Guwahati, India. 

\end{document}